\documentclass[slac_one]{revtex4}
\usepackage{graphicx}
\usepackage{fancyhdr}
\begin{document}

\title{Tracking Performance of the ATLAS Inner Detector \\ and Observation of Known Hadrons} 

%

\author{Manuel Kayl (mkayl@nikhef.nl), \\ on behalf of the ATLAS collaboration}
\affiliation{Nikhef-Nationaal instituut voor subatomaire fysica,\\
 Science Park 197, 1098 XG Amsterdam, Netherlands}

\begin{abstract}
The inner detector is the central tracking device of the ATLAS detector. In these proceedings the tracking performance of the inner detector is presented on collision data recorded at $\sqrt{s}$ = 900 GeV and 7 TeV. The identification of resonances like  $\Xi$ and $\Omega$ baryons in cascade decays via $K_{\mathrm{s}}$ and $\Lambda$ mesons is presented as well as the reconstruction of the $J / \psi $ and  $\psi$(2S) mesons decaying into two muons. Furthermore, the performance of the track reconstruction and a data-driven method of estimating the track reconstruction efficiency as used in measurements of charged particle densities are discussed.
\end{abstract}

\maketitle

\thispagestyle{fancy}


\section{The ATLAS Inner Detector} 

\begin{figure}[htbp]
\centering
\includegraphics[width=0.7\textwidth]{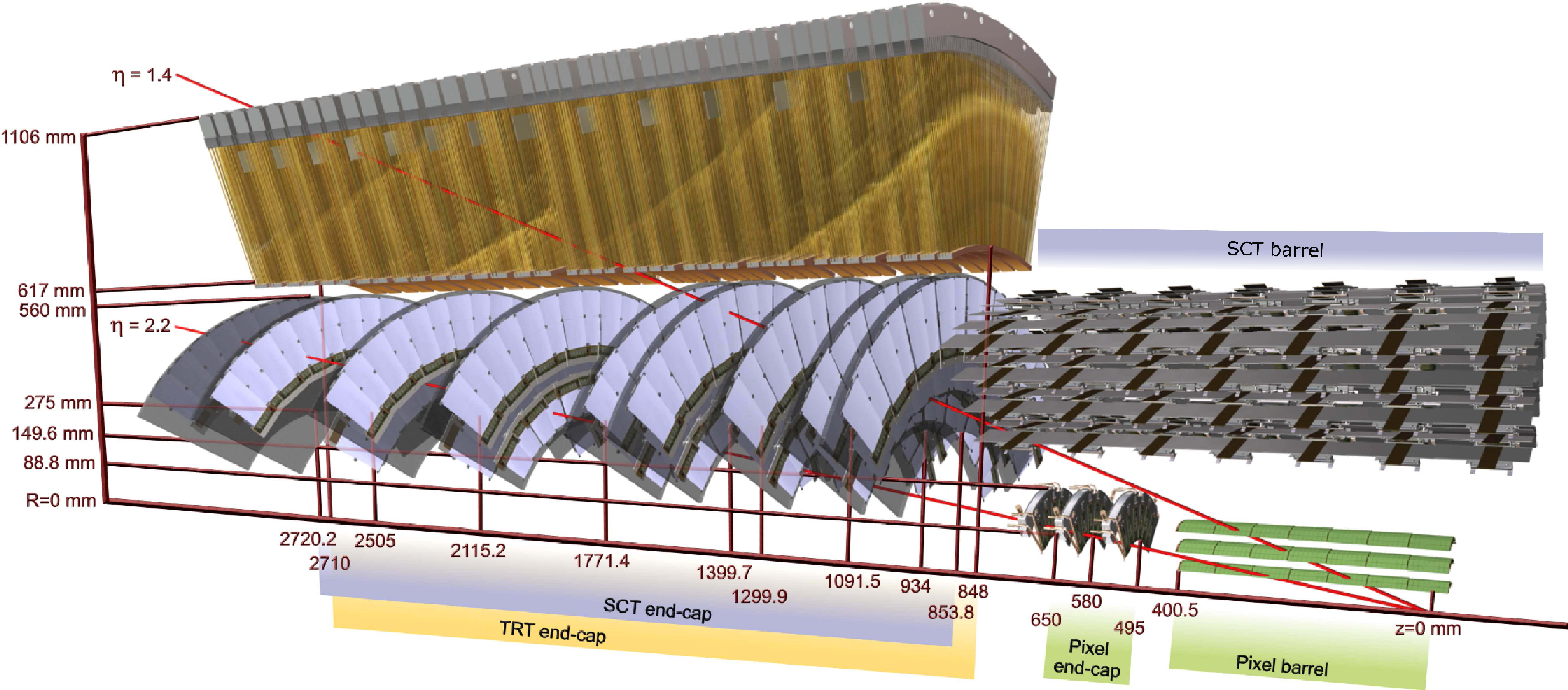} 
\caption{Drawing of the ATLAS Inner Detector traversed by two 10 GeV  tracks with $~ \mathrm{\eta} = 1.4$ and $2.2$. The TRT barrel detector is not shown.}
\label{fig:ID}
\end{figure}

The inner detector  is the innermost tracking device of ATLAS located in a $2 ~ \mathrm{T}$ solenoid magnet~\cite{id_tdr}. The inner detector consists of a silicon pixel detector, a silicon strip detector (SCT) and a straw tube transition radiation tracker (TRT). Its main design goals are the primary and secondary vertex identification as well as a precise reconstruction of charged particles in the region $| \mathrm{\eta}|  < 2.5$. A cutaway view of the inner detector is shown in figure~\ref{fig:ID}. Hit resolutions in the barrel of 10 $\mathrm{\mu m}$ ($\mathrm{R \phi}$) x 115 $\mathrm{\mu m}$ ($\mathrm{z}$) in the pixel, 17 $\mathrm{\mu m}$ ($\mathrm{R \phi}$) x 580 $\mathrm{\mu m}$ ($\mathrm{z}$) for the SCT and 130 $\mathrm{\mu m}$ ($\mathrm{R \phi}$) in the TRT detectors provide the high granularity needed to meet the design requirements~\cite{detPaper}. 

\vspace{-1 cm}

\section{Tracking performance of the inner detector}

\begin{figure}[htbp]
\centering
\begin{minipage}[t]{0.4\linewidth}
    \vspace{0.8 cm}
    \includegraphics[width=1\textwidth]{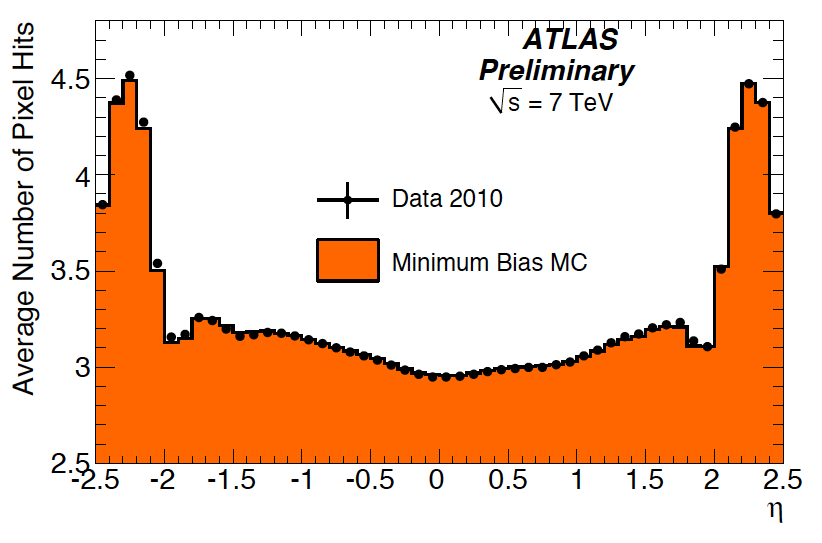} 
\end{minipage}
\begin{minipage}[t]{0.4\linewidth}
    \vspace{0.1 cm}
\includegraphics[width=1\textwidth]{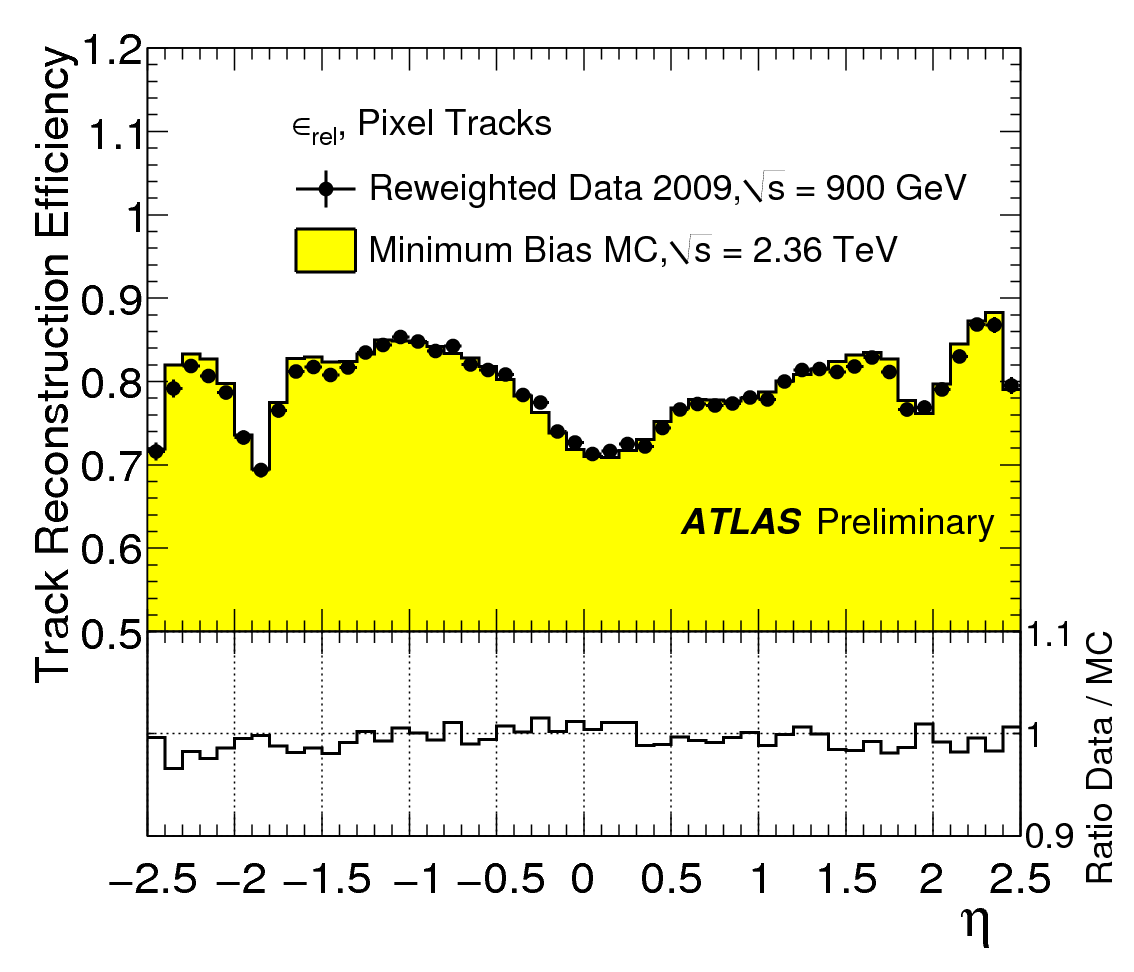}
\end{minipage}

\caption{Average number of pixel hits as a function of $\eta$ (left). Relative track reconstruction efficiency to reconstruct a track in the pixel detector if a track was found in the SCT and TRT detectors (right).}
\label{fig:perf}
\end{figure}

The performance of the inner detector track reconstruction is illustrated in figure~\ref{fig:perf} (left), where the average number of pixel hits per track is shown as a function of $\eta$~\cite{MB15}. The agreement between collision data and simulation indicates the excellent understanding of the detector geometry. Another important measure for the performance of the detector is the efficiency to reconstruct charged particles. In general the track reconstruction efficiency has to be obtained from simulation which is subject to sizable uncertainties arising for example from the description of the amount of material. A method has been developed to reduce the uncertainty by computing the efficiency to reconstruct a track solely from hits in the pixel detector if a track was already found in the SCT and TRT detectors~\cite{MB236}. A comparison between data and simulation of this relative track efficiency reconstruction efficiency, which is sensitive to improper descriptions of the amount of material in the inner detector, is shown in figure~\ref{fig:perf} (right). The agreement between data and simulation underlines the good understanding of the track reconstruction.

\section{Observation of Hadron Resonances}

A precise and efficient identification of mesons and baryons produced in proton-proton collision is a vital ingredient for many heavy flavour analyses and probes the combined performance of track and vertex reconstruction algorithms of the inner detector. 

\subsection{Reconstruction of Cascade Decays}

\begin{figure}[htb!]
\begin{center}
\begin{minipage}[t]{0.25\linewidth}
    \vspace{0.1 cm}
    \includegraphics[width=1\textwidth]{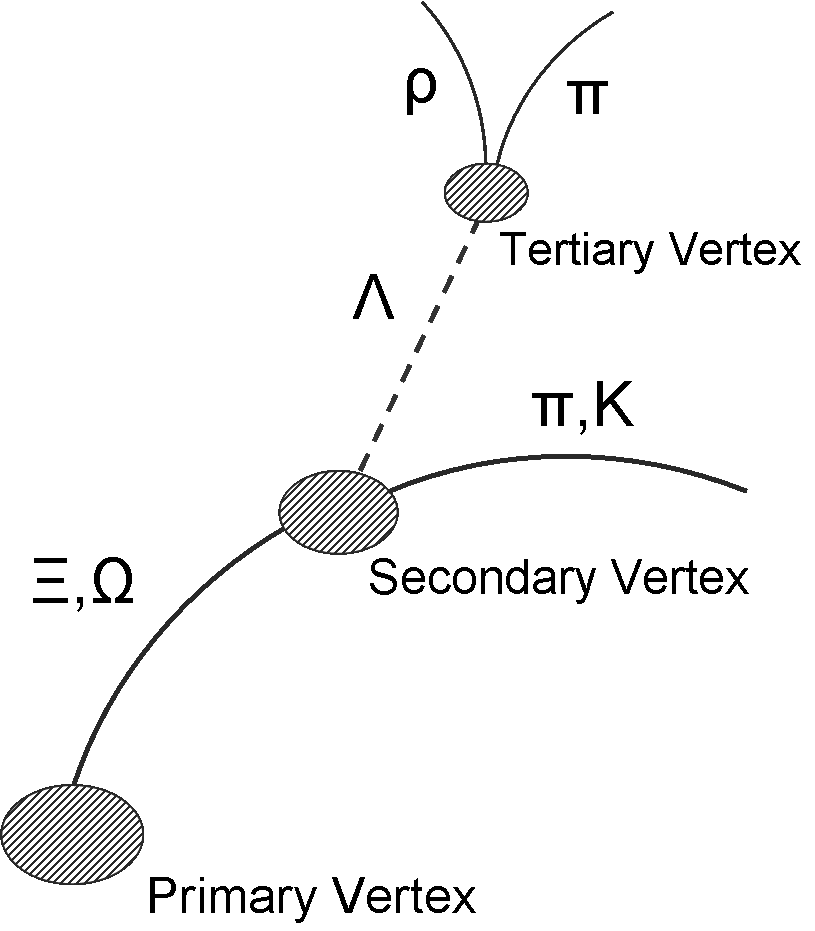} 
\end{minipage}
\begin{minipage}[t]{0.4\linewidth}
    \vspace{0.01 cm}
\includegraphics[width=1\textwidth]{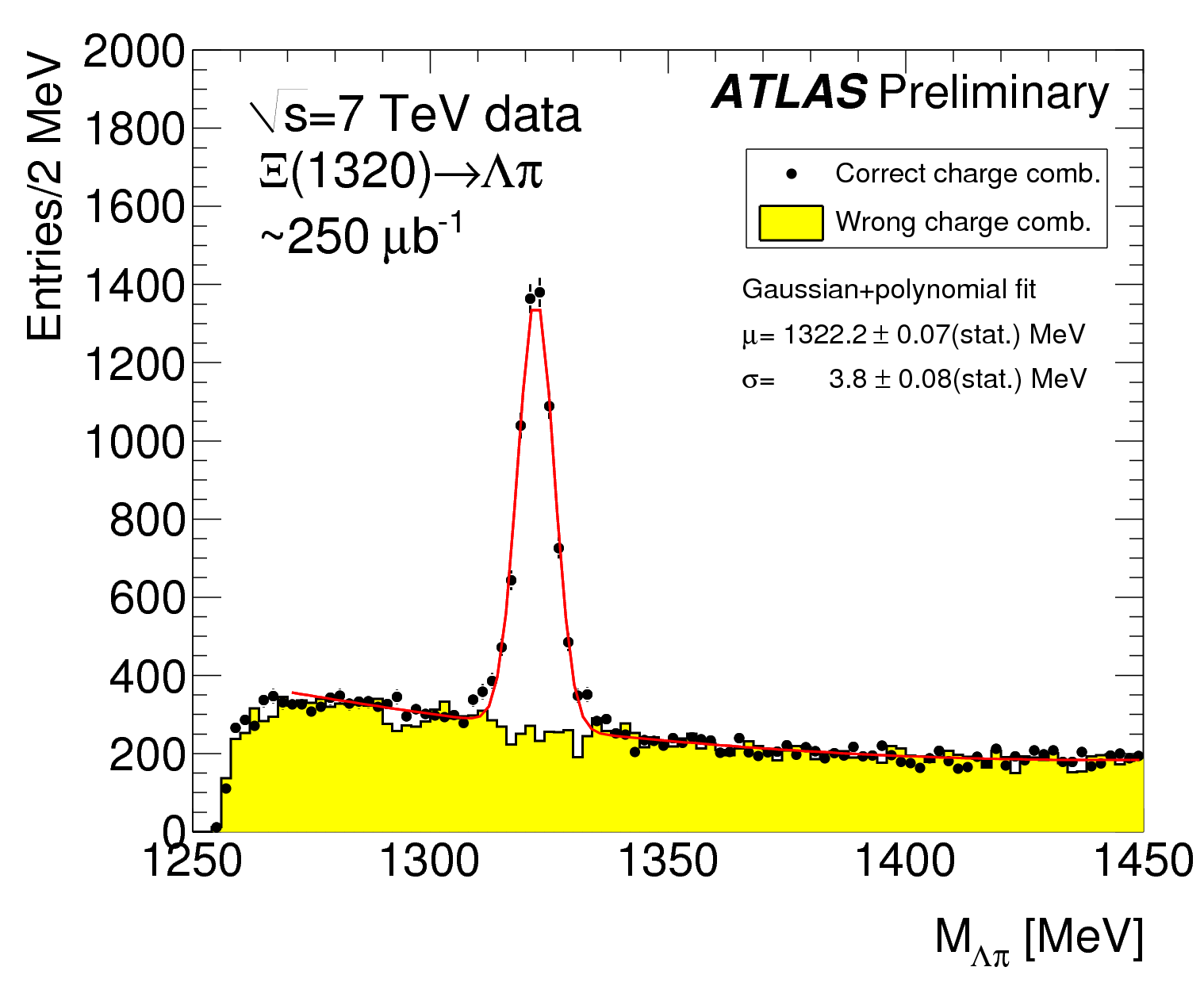}
\end{minipage}
\caption{Schematic view of the decay of the $\Xi$ and $\Omega$ resonances (left). Invariant mass distribution of $\Xi$ distribution (right).\label{fig:Xi}}
\end{center}
\end{figure}

The $\Xi$ and $\Omega$ resonances have decay channels with long-lived particles $K_{\mathrm{s}}$ and $\Lambda$ in the final state which also decay inside the volume of the inner detector. The hyperons are observed in the decay modes $\Xi \rightarrow \pi \Lambda(p \pi)$ and $\Omega \rightarrow K \Lambda(p \pi)$~\cite{xi}. Both hyperon charged states are used in analysis. Due to their significant lifetimes both $\Xi$ and $\Omega$ decay far from the primary event vertex producing secondary vertices. The subsequent decay of the $\Lambda$ produces a tertiary vertex as illustrated in figure~\ref{fig:Xi} (left). A specialized cascade reconstruction algorithm, which is able to fit the secondary and tertiary vertices simultaneously while constraining the intermediate particles to point to their production vertices, was used to reconstruct the hyperons. In figure~\ref{fig:Xi} (right) the invariant mass distribution of the 
$\Xi$ candidates is shown together with an estimation of the background. The background was estimated from data by using the same reconstruction algorithm but requiring a wrong charge combination of pions and protons. The fitted mass of the $\Xi$ hyperon is (1322.2 $\pm$ 0.07 (stat)) MeV whereas the value from \emph{The Review of Particle Physics} is 1321.7 MeV~\cite{PDG}. The precision of the fitted mass is currently limited by the the knowledge of the alignment and the magnetic field.

\vspace{-0.5 cm}

\subsection{Reconstruction of the $J / \psi $ and  $\psi$(2S) mesons}

\begin{figure}[htb!]
\begin{center}
\begin{minipage}[t]{0.4\linewidth}
    \includegraphics[width=1\textwidth]{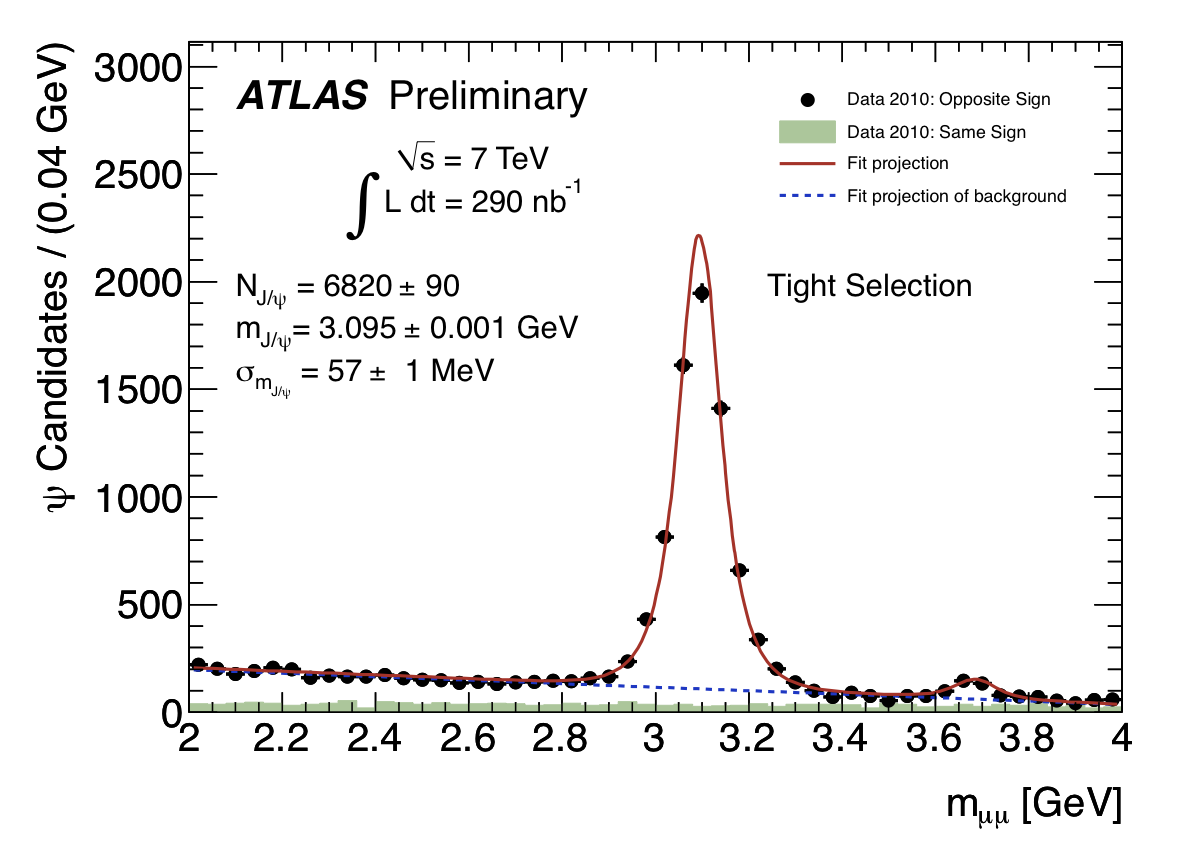} 
\end{minipage}
\begin{minipage}[t]{0.4\linewidth}
\includegraphics[width=1\textwidth]{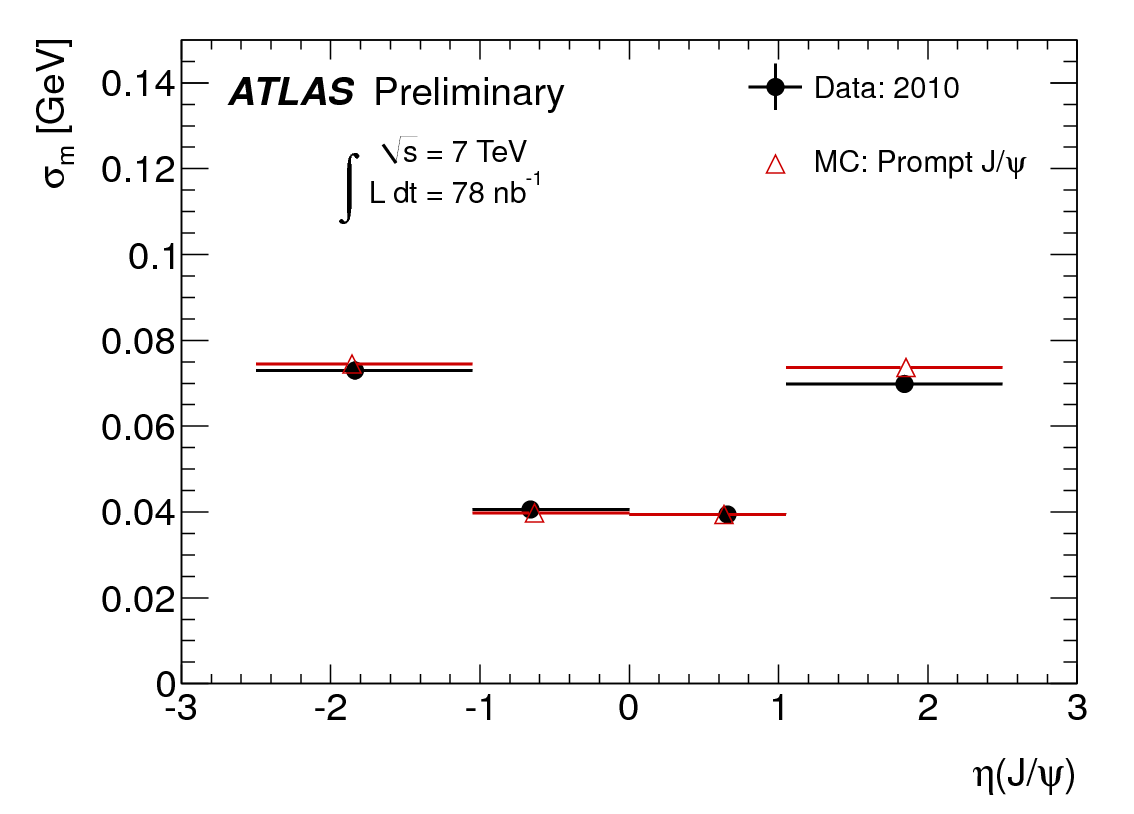}
\end{minipage}
\caption{Di-muon invariant mass spectrum showing peaks corresponding to the $J / \psi $ and  $\psi$(2S) mesons (left). Mass resolution as a function of $\eta$ of the $J / \psi $ candidate (right). \label{fig:Jpsi}}
\end{center}
\end{figure}

Whereas the studies presented so far mainly investigated the performance on kaons and pions, the reconstruction of the $J / \psi $ and  $\psi$(2S) mesons decaying into two muons investigates the performance of the muon reconstruction of the inner detector~\cite{jpsi}. Tracks were identified as muons by looking for coincident signals in the inner detector and the muon spectrometer. The track parameters used for the analysis are derived using only information from the inner detector. Two peaks arise in the invariant mass spectrum of two oppositely charged muons as shown in figure~\ref{fig:Jpsi} (left) corresponding to the $J / \psi $ and  $\psi$(2S) mesons. The mass and width are obtained by fitting a Gaussian with a second-order polynomial to model the background to the invariant mass distribution. The measured central value is (3.095 $\pm$ 0.001 (stat)) GeV for the $J / \psi $ meson compared to 3.097 GeV  from \emph{The Review of Particle Physics}.  

In figure~\ref{fig:Jpsi} (right) a comparison of the mass resolution of the $J / \psi $ as a function of the pseudorapidity of the $J / \psi $ between data and simulation is shown. As predicted by the simulation, the mass resolution of approximately 35 MeV in the central part of the detector is a factor of two better than in the forward region of the detector. This is due to the fact that more material is present at high pseudorapidity and muons undergo more multiple scattering which worsens the resolution of the track parameters.

\vspace{-1 cm}

\section{Conclusions}

The performance of the ATLAS inner detector on proton-proton collision data was presented in these proceedings. Studies which mainly deal with the  reconstruction of pions and kaons such as the estimation of the track reconstruction efficiency and the identification of cascade decays were presented. Additionally, the performance of the muon reconstruction was investigated by analysing the decay of the $J / \psi $ and  $\psi$(2S) mesons into muons.  The agreement reached with the predictions from Monte Carlo simulation and with the mass values from \emph{The Review of Particle Physics} is consistent with the current precision of the knowledge of the magnetic field as well as the alignment and underlines the good understanding of the performance of the inner detector.



\end{document}